\documentclass[aps,prb,twocolumn,groupedaddress,floatfix]{revtex4-1}

\usepackage{graphicx}
\usepackage{dcolumn}
\usepackage{bm}

\begin{document}

\title{Gate control of  Berry phase  in III-V semiconductor quantum dots }
\author{Sanjay Prabhakar,$^{1}$    Roderick Melnik$^{1}$   and  Luis L Bonilla,$^{2}$  }
\affiliation{
$^1$ The MS$\mathrm{2}$Discovery Interdisciplinary Research Institute, M\,$^2$NeT Laboratory, Wilfrid Laurier University, Waterloo, ON, N2L 3C5 Canada\\
$^2$Gregorio Millan Institute, Universidad Carlos III de Madrid, 28911, Leganes, Spain
}

\date{May 14, 2014}

\begin{abstract}
We analyze the  Berry phase in III-V semiconductor quantum dots (QDs).   We show that the Berry phase is highly sensitive to    electric fields arising from the interplay between the Rashba and  Dresselhaus spin-orbit (SO) couplings.
We report that the accumulated Berry phase can be induced from  other available quantum states that  differ only by one quantum number of the corresponding spin state. The  sign change in  the $g$-factor due to the penetration of Bloch wavefunctions into the barrier materials  can be reflected   in the Berry phase. We provide characteristics of the Berry phase for three different length scales (spin-orbit length, hybrid orbital length and orbital radius). We solve the time dependent Schr$\mathrm{\ddot{o}}$dinger equation by utilizing the Feynman disentangling technique, and we investigate the evolution of  spin dynamics during the adiabatic transport of  QDs in the two-dimensional plane. Our results can pave the way to  building a topological   quantum computer in which the Berry phase can be engineered and be manipulated with the application of the spin-orbit couplings through gate-controlled electric fields.
\end{abstract}

\maketitle

\section{Introduction}
Coherent control of single-electron spin relaxation and its measurement in III-V semiconductor quantum dots  might provide a new foundation of architecture  to build  next-generation spintronic devices.~\cite{amasha08,takahashi10,prabhakar13,awschalom02,loss98,prabhakar09,prabhakar11}
To manipulate the spin in quantum dots (QDs), achieving the state of the art of semiconductor technology, a number of researchers have recently proposed measuring spin behavior  in QDs by utilizing  electric fields generated by isotropic and anisotropic  gate potentials.~\cite{amasha08,takahashi10,sousa03,prabhakar09,prabhakar11,prabhakar12}
The effective $g$-factor and phonon mediated  spin relaxation in both isotropic and anisotropic QDs  can be tuned with spin-orbit coupling.~\cite{sousa03,takahashi10,prabhakar12}  The electric-field control of  spin  provides an opportunity  for tuning the spin
current on and off in QDs formed in a single electron transistor. Such control can help to initialize the electron spin  in spintronic devices.~\cite{amasha08,takahashi10,bandyopadhyay00,flatte11}

Alternatively, a more robust technique can be applied to manipulate    single electron spins in QDs  through the non-Abelean geometric phases.~\cite{giuliano03,aleiner01,wang08,hwang06,yang06,yang07,berry84} For a system of degenerate quantum states, Wilczek and Zee showed that the geometric
phase factor is replaced by a non-Abelian time dependent unitary operator acting on the initial states within the subspace of degeneracy.~\cite{wilczek84,prabhakar10} Since then,  the geometric phase has been measured  experimentally for a variety of  systems, such as quantum states driven by a microwave field~\cite{pechal12} and  qubits with tilted magnetic fields.~\cite{berger12,leek07}
Manipulation of spin qubits through the Berry phase implies  that   injected data can be read out with a different phase that is  topologically protected from the outside world.~\cite{sarma05,hou00,loss90,yaroslav11,jose06,jose06a,jose07,jones00,falci00}
Several recent reviews of the  Berry phase have been presented  in Refs.~\onlinecite{xiao10,nayak08}.
One of the promising  research proposals for building a solid-state topological quantum computer
is  that the accumulated Berry phase in a QD system  may be manipulated using the interplay between  the Rashba-Dresselhaus spin-orbit couplings.~\cite{jose06,aleiner01}
The Rashba spin-orbit coupling arises from the asymmetric triangular quantum well along the growth direction, while the Dresselhaus spin-orbit coupling arises due to bulk inversion asymmetry in the crystal lattice.~\cite{bychkov84,dresselhaus55}
A recent work by Bason et al. shows that the Berry phase can be measured for a two level quantum system in a superadiabatic basis comprising the Bose-Einstein condensates in optical lattices.~\cite{bason12}

Recently, it has also been shown that the geometric phase can be induced  on the electron spin states in QDs by moving  the dots adiabatically in a closed loop in the two dimensional  ($2$D) plane  plane through  application of a gate controlled electric field.~\cite{prabhakar10,jose06a,ban12,prabhakar14} Furthermore, the authors in Refs.~\onlinecite{bednarek12,bednarek08,bednarek08a} have recently proposed  building a QD device in the absence of magnetic fields that can perform  quantum gate operations (NOT gate, Hadamard gate and Phase gate) using an externally applied  sinusoidal varying potential through external gates.

In this paper, we show how to transport electron spin states of QDs in the presence of externally applied magnetic fields along the z-direction    in a closed loop  through the application of a time dependent distortion potential. We investigate  the interplay between the Rashba and  the Dreeselhaus spin-orbit couplings on the scalar Berry phase.~\cite{hwang06,wu11} The transport of the dots is carried out very slowly, so that the adiabatic theorem can be applied on the evolution of the spin dynamics. In particular,  the sign change in the $g$-factor of electrons in the QDs due to the penetration of the Bloch wavefunctions into the barrier materials can be  manipulated with the interplay between the Rashba-Dresselhaus spin-orbit couplings in the Berry phase. We show that the Berry phase in QDs can be engineered and therefore manipulated with the application of  spin-orbit couplings through gate controlled electric fields. We solve the time dependent Schr$\mathrm{\ddot{o}}$dinger equation and investigate the evolution of spin dynamics in QDs.

The paper is organized as follows. In Sec.~\ref{theoretical-model}, we provide a detailed theoretical formulation of the total Hamiltonian of a moving QD in relative coordinates and relative momentum. We also show that the quasi adiabatic variables are gauged away from the total Hamiltonian. Here we write the total Hamiltonian of the moving dots in terms of annihilation and creation operators and we utilize the perturbation theory to find the analytical expression of the Berry phase. In Sec.~\ref{computational-method}, we provide details of our computational methodology.  In Sec.~\ref{results-discussion}, we investigate the interplay between the Rashba-Dresselhaus spin-orbit couplings on the Berry phase of III-V semiconductor  QDs. Finally, in Sec.~\ref{conclusion}, we summarize our results.

\section{Theoretical Model} \label{theoretical-model}

We consider the Hamiltonian $H = H_{0} +  H_{so}$ for an electron in a QD of the III-V semiconductor.~\cite{prabhakar09,prabhakar11} Here
\begin{equation}
H_{0} = {\frac {\left\{\mathbf{p}+e\mathbf{A}(\mathbf{r})\right\}^2}{2m}} + {\frac{1}{2}} m \omega_o^2\mathbf{r}^2 + e\mathbf{E}(t)\cdot \mathbf{r} + \frac{g_0 \mu B \sigma_z}{2}
\label{hxy}
\end{equation}
is the Hamiltonian for a QD electron in the x-y plane of the two-dimensional electron gas (2DEG) in the presence of a uniform magnetic field B along the z-direction and a time dependent lateral electric field $\mathbf{E}(t)$. The second term is the spin-orbit Hamiltonian consisting of the Rashba and the linear Dresselhaus couplings, $H_{so}=H_R+H_D$, where
\begin{eqnarray}
H_R =\frac{\alpha_R}{\hbar}\left\{\sigma_x \left(p_y+eA_y\right) - \sigma_y \left(p_x+eA_x\right)\right\},\label{rashba}\\
H_D= \frac{\alpha_D}{\hbar}\left\{-\sigma_x \left(p_x+eA_x\right) + \sigma_y \left(p_y+eA_y\right)\right\}, \label{dresselhaus}
\end{eqnarray}
with $\alpha_R=\gamma_ReE_z$ and $\alpha_D=0.78\gamma_D\left(2me/\hbar^2\right)^{2/3}E_z^{2/3}$.

In~(\ref{hxy}), $\mathbf{r}=\left(x,y,0\right)$ is the position vector and $\mathbf{p} = -i\hbar (\partial_x,\partial_y,0)$ is  the canonical momentum. The vector potential $\mathbf{A}(\mathbf{r})$ is due to the applied magnetic field $\mathbf{B}$. Here $-e <  0$ is the electronic charge, $m$ is the effective mass of an electron and $\mu$ is the Bohr magneton. The confining potential is  parabolic with the center at  $\mathbf{r}=0$. The third term in~(\ref{hxy}) is the electric potential energy due to an applied periodic lateral electric field $\mathbf{E}(t)=\left(E_x(t),E_y(t),0\right)$, where $E_x(t)=E_0\cos\omega t$ and $E_y(t)=E_0\sin\omega t$. Varying $\mathbf{E}(t)$ very slowly, we treat its two components as adiabatic parameters. In principle,  the alternating  electric field induces a vector potential added to the one due to the applied  uniform magnetic field $\mathbf{B}$ in the z-direction. However, such a contribution to  $\mathbf{B}$ is in practice extremely small as reported earlier by Golovach et. al.~\cite{golovach06}  Our estimate shows that by using  the dot size $\ell_0=20~\mathrm{nm}$, the orbital radius $r_0\approx 70\mathrm{nm}$, the frequency  $\omega=1~\mathrm{THz}$, and the maximum electric field $E_0=0.5~\mathrm{mV/nm}$,  the magnitude of the induced magnetic field is approximately  $B_{in} \approx \epsilon_r\mu_r\pi r_0\omega E_0/(2c^2)\approx \epsilon_r\mu_r \times10^{-6}~tesla$, where $\epsilon_r$ and $\mu_r$ are the relative electric permittivity and the magnetic permeability, respectively (for mathematical derivation, see appendix~\ref{appendix-A}). This contribution is negligible  compared to the applied $\mathrm{\mathbf{B}}$ field.
Therefore, for the vector potential, we simply choose the gauge of the form $A(r)=B/2\left(-y,x,0\right)$. The last term of~(\ref{hxy}) describes the Zeeman coupling with $g_0$, the bulk $g$-factor. Saniz et. al.~\cite{saniz03} have suggested that the Coulomb repulsion between electrons with opposite spins of strongly correlated systems would give rise to appreciable oscillations in spin polarization. For weakly correlated systems, such effect vanishes.  Hence, the Berry phase in QDs, for strongly correlated systems, is also influenced by Coulomb repulsion. As is pointed out by Saniz \emph{et. al.},~\cite{saniz03} the Coulomb coupling becomes weaker with decreasing electron density and increasing dot size. Since the dot size of our choice is $\ell_0=20~\mathrm{nm}$, the Coulomb coupling is very small as compared with the Zeeman coupling. Thus, in our model, the Coulomb coupling is not included. For  strongly correlated systems with $\ell_0=0.5~\mathrm{nm}$ or less, the Coulomb coupling can not be ignored.

The electric field at a fixed time $t_0$ effectively shifts   the center of the parabolic potential from $\mathbf{r}=0$ to $\mathbf{r}=\mathbf{r_0} \left(t_0\right)$, where $\mathbf{r}_0=-e\mathbf{E}\left(t_0\right)/m\omega_0^2$. Hence the Hamiltonian~(\ref{hxy})  can be expressed as
\begin{equation}
H_{0} = {\frac {\left\{\mathbf{p}+e\mathbf{A}(\mathbf{r})\right\}^2}{2m}} + {\frac{1}{2}} m \omega_o^2\left(\mathbf{r}-\mathbf{r_0}\right)^2 -G + {\frac \Delta 2}\sigma_z,
\label{hxy-1}
\end{equation}
where  $G=e^2E_0^2/\left(2m\omega^2_0\right)$ is an unimportant constant and $\Delta=g_0\mu B$ is the Zeeman energy.
As the applied $\mathbf{E}-$field varies, the QD will be adiabatically transported along a circle of radius $r_0=|\mathbf{r}_0|=eE_0/m\omega^2_0$.

At this point,  we introduce the relative coordinate $\mathbf{R}=\mathbf{r}-\mathbf{r_0}=\left(X,Y,0\right)$ and  the relative momentum $\mathbf{P}=\mathbf{p}-\mathbf{p_0}=\left(P_X,P_Y,0\right)$, where $\mathbf{p_0}$ is the momentum of the slowly moving dot which may be classically given by $m\mathbf{\dot{r}}_0$. Obviously $\left[X,P_X\right]=\left[Y,P_Y\right]=i\hbar$ and $\left[X,P_Y\right]=\left[Y,P_X\right]=0$. We can show that the adiabatic variables $\mathbf{p}_0$ and $\mathbf{r}_0$ will be gauged away from the Hamiltonian by the transformation $\tilde{H}=UHU^{-1}$ and $\tilde{\psi}=U\psi$ with $U=\exp\left\{ \left(i/\hbar\right) \left(\mathbf{p_0}+e\mathbf{A}\left(\mathbf{r}_0\right)\right)\cdot \mathbf{R}\right\}$, so that
\begin{eqnarray}
\tilde{H}_{0} = {\frac {1}{2m}}\left\{\mathbf{P}+e\mathbf{A}(\mathbf{R})\right\}^2 + {\frac{1}{2}} m \omega_o^2R^2 -G+ {\frac \Delta 2}  \sigma_z,\label{tilde-hxy}\\
\tilde{H}_{so} =UH_{so}\left(\mathbf{p},\mathbf{r}\right)U^{-1}=H_{so}\left(\mathbf{P},\mathbf{R}\right),\label{Hso}
\end{eqnarray}
where $\mathbf{A}\left(\mathbf{P},\mathbf{R}\right)=\left(B/2\right)\left(-Y,X,0\right)$. This means that the electron in the shifted dot obeys a quasi-static eigenequation, $\tilde{H}\left(\mathbf{P},\mathbf{R}\right)\tilde{\psi}_n\left(\mathbf{R}\right)=
\tilde{\varepsilon}_n\tilde{\psi}_n\left(\mathbf{R}\right)$, where $\tilde{H}=\tilde{H}_0+\tilde{H}_{so}$. By an adiabatic transport of the dot, the eigenfunction $\tilde{\psi}_n$ will acquire the Berry phase as well as the usual dynamical phase. Namely, $\psi_n\left(\mathbf{r},t\right)=e^{i\gamma_n\left(t\right)} e^{i\theta_n\left(t\right)}U^{-1}\tilde{\psi}_n\left(R\right)$, where $\gamma_n$ is the Berry phase and $\theta_n$ is the dynamical phase.

In order to evaluate the Berry phase explicitly, we return to the original Hamiltonian $H$ and put it in the form:
\begin{equation}
H=\tilde{H}_0\left(\mathbf{P},\mathbf{R}\right)+H_{so}\left(\mathbf{P},\mathbf{R}\right)+
H_{ad}\left(\mathbf{P},\mathbf{R};\mathbf{p_0},\mathbf{r_0}\right),
\label{total-adiabatic}
\end{equation}
where
\begin{eqnarray}
H_{ad}\left(\mathbf{P},\mathbf{R};\mathbf{p}_0,\mathbf{r}_0\right)&=&\frac{1}{m}\left\{\mathbf{P}
+e\mathbf{A}\left(\mathbf{R}\right)\right\}\cdot \left\{\mathbf{P}+e\mathbf{A}\left(\mathbf{R}\right)\right\}\nonumber\\
&&+H_{so}\left(\mathbf{p}_0,\mathbf{r}_0\right)+G',\label{Had}
\end{eqnarray}
with another unimportant constant $G'=\left\{\mathbf{p}_0+e\mathbf{A}\left(\mathbf{r}_0\right)\right\}^2/\left(2m\right)^2=\left(\omega+\omega_c/2\right)^2 r_0^2/4$.

The quasi-static Hamiltonian $\tilde{H}_0\left(\mathbf{P},\mathbf{R}\right)$ can be diagonalized on the basis of the number states $|n_+,n_-,\pm 1\rangle$:
\begin{equation}
\tilde{H}_0=\left(N_+ + \frac{1}{2}\right)\hbar\Omega_+ + \left(N_- + \frac{1}{2}\right)\hbar\Omega_- -G + \frac{\Delta}{2}\sigma_z,\label{H0}
\end{equation}
where $N_{\pm}=a^\dagger_{\pm}a_{\pm}$ are the number operators with eigenvalues $n_\pm \in N_0$. Here,
\begin{eqnarray}
a_{\pm}=\frac{1}{\sqrt{4m\hbar\Omega}}\left(iP_x \pm P_y\right)+\sqrt{\frac{m\Omega}{4\hbar}}\left( X \mp i Y \right),\\ \label{a-lowering}
a_{\pm}^\dagger=\frac{1}{\sqrt{4m\hbar\Omega}}\left(-iP_x \pm P_y\right)+\sqrt{\frac{m\Omega}{4\hbar}}\left( X \pm i Y \right),\label{a-raising}
\end{eqnarray}
provided that $\left[a_{\pm},a_{\pm}^\dagger\right]=1$.
Correspondingly, the other terms may also be expressed in terms of the raising and lowering operators,
\begin{eqnarray}
H_{so}\left(\mathbf{P},\mathbf{R}\right)&=&\alpha_R\left(\xi_+\sigma_+ a_+ - \xi_- \sigma_- a_-\right)~~~~~~\nonumber\\
&&+i\alpha_D\left(\xi_+\sigma_- a_+ + \xi_- \sigma_+ a_-\right)+H.c.,~~~~~~\label{Hso-1}\\
H_{ad}&=&\frac{\hbar}{2}\left(\xi_+z_+a_+-\xi_-z_-a_-\right)\omega_+ \nonumber\\
&&+\frac{1}{\hbar}\left(\alpha_Rz_--i\alpha_Dz_+\right)m\omega_+\sigma_+ +H.c. \label{had}
\end{eqnarray}
In the above, we have used the notations, $z_{\pm}=x_0\pm iy_0$, $\xi_{\pm}= \sqrt {m\Omega/\hbar}\pm eB/\sqrt{4m\hbar\Omega}$, $\sigma_{\pm}=\left(\sigma_x\pm i\sigma_y\right) /2$, $\omega_{\pm}=\omega\left( 1 \pm \omega_c/\left(2\omega\right)\right)$, $\Omega_{\pm}=\Omega \pm \omega_c/2$ and $\Omega=\sqrt {\omega_0^2+\omega_c^2/4}$ with $\omega_c=eB/m$  being the cyclotron frequency. In~(\ref{Hso-1}) and~(\ref{had}), $H.c.$ signifies the Hermitian conjugate.

For III-V semiconductor QDs, we define the SO lengths  $\lambda_R=\hbar^2/m\alpha_R$ and $\lambda_D=\hbar^2/m\alpha_D$ and estimate that  the SO lengths  are much larger than the hybrid orbital length $\ell$  and  QDs radius $\ell_0$ (see left panel of Fig.~\ref{fig4}). Therefore the Rashba-Dresselhaus SO couplings Hamiltonians are considered as a small perturbations.
Based on the second order perturbation theory, the four lowest energy eigenvalues  of the moving dot  are given by
\begin{widetext}
\begin{eqnarray}
\varepsilon_{0,0,-1} = \hbar\Omega-G-\frac{\Delta}{2} -\frac{\alpha_R^2\xi_-^2}{\hbar\left(\Omega-\omega_c/2\right)+\Delta}  -\frac{\alpha_D^2\xi_+^2}{\hbar\left(\Omega+\omega_c/2\right)+\Delta} + \varepsilon^{(2)}_{0,0,-1},~~~~~~  \label{varepsilon-00d}\\
\varepsilon_{0,0,1} = \hbar\Omega-G+\frac{\Delta}{2} -\frac{\alpha_R^2\xi_+^2}{\hbar\left(\Omega+\omega_c/2\right)-\Delta}  -\frac{\alpha_D^2\xi_-^2}{\hbar\left(\Omega-\omega_c/2\right)-\Delta} + \varepsilon^{(2)}_{0,0,1}, ~~~ ~~~ \label{varepsilon-00u}\\
\varepsilon_{0,1,-1} = \frac{1}{2} \hbar\left(\Omega_++3\Omega_-\right)-G-\frac{\Delta}{2} -\frac{2\alpha_R^2\xi_-^2}{\hbar\left(\Omega-\omega_c/2\right)+\Delta}  -\frac{\alpha_D^2\xi_+^2}{\hbar\left(\Omega+\omega_c/2\right)+\Delta} + \frac{\alpha_D^2\xi_-^2}{\hbar\left(\Omega-\omega_c/2\right)-\Delta}+\varepsilon^{(2)}_{0,1,-1}, ~~~~~ \label{varepsilon-01d}\\
\varepsilon_{0,1,1} = \frac{1}{2} \hbar\left(\Omega_++3\Omega_-\right)-G+\frac{\Delta}{2} -\frac{2\alpha_D^2\xi_-^2}{\hbar\left(\Omega-\omega_c/2\right)-\Delta}  +\frac{\alpha_R^2\xi_-^2}{\hbar\left(\Omega-\omega_c/2\right)+\Delta} - \frac{\alpha_R^2\xi_+^2}{\hbar\left(\Omega+\omega_c/2\right)-\Delta} +
\varepsilon^{(2)}_{0,1,1},~~~~~ \label{varepsilon-01d}
\end{eqnarray}
\end{widetext}
where
\begin{eqnarray}
\varepsilon^{(2)}_{00\pm 1}&=&\pm\left(\frac{m\omega_+}{\hbar\sqrt\Delta}\right)^2\{(\alpha^2_R+\alpha^2_D)r_0^2-4\alpha_R\alpha_D x_0 y_0\}\nonumber\\
&&-\left(\frac{ \hbar\omega^2_+}{4}\right) \left\{\frac{\xi^2_-}{\Omega_-}+\frac{\xi^2_+}{\Omega_+}\right\}r_0^2=\varepsilon^{(2)}_{01\pm 1}.\label{varepsilon-ad}
\end{eqnarray}
Since $x_0y_0 = r_0^2\sin2\theta/2 $, we conclude that the energy spectrum of the dot  depends on the rotation angle. As a result,
it is possible to have the interplay between the spin-orbit coupling and the evolution of spin dynamics during the adiabatic transport of the dots (see Fig.~\ref{fig2}).
In the above equation, we see that $\varepsilon^{(2)}_{00\pm 1 }=\varepsilon^{(2)}_{01\pm 1}$.  This means that the Berry phase  depends not on
how quantum states of the dot traveled but only  on the total adiabatic area enclosed  during the adiabatic transport of the dot in the 2D plane.
\begin{figure}
\includegraphics[width=8cm,height=7.5cm]{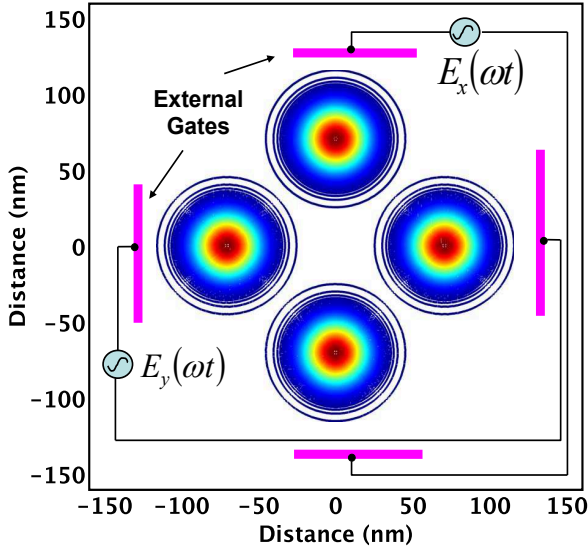}
\caption{\label{fig1} (Color online) Contour plots of the realistic electron wave functions in GaAs QDs that are adiabatically transported in one complete rotation in the plane of 2DEG  under the influence of externally applied gate potential. We chose $E_0=5\times 10^3$ V/cm,  $E_z=5 \times 10^5$V/cm, $B=1$T and  $\ell_0=20$nm. Here we report that $\varepsilon_{0,1,+1}-\varepsilon_{0,0,+1}\approx 2.1$ meV  which is constant during the adiabatic transport  of the QDs.  }
\end{figure}
\begin{figure}
\includegraphics[width=8.5cm,height=4.5cm]{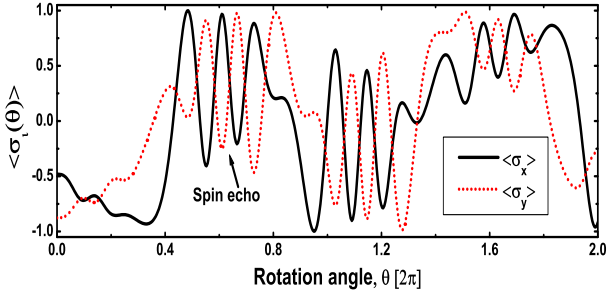}
\caption{\label{fig2} (Color online) Evolution of spin dynamics during the adiabatic transport of the GaAs QDs (see appendix~\ref{appendix-B}).  The  parameters  are chosen the same as in Fig.~\ref{fig1} but  $\ell_0=30$nm.   }
\end{figure}
\begin{figure*}
\includegraphics[width=18cm,height=8cm]{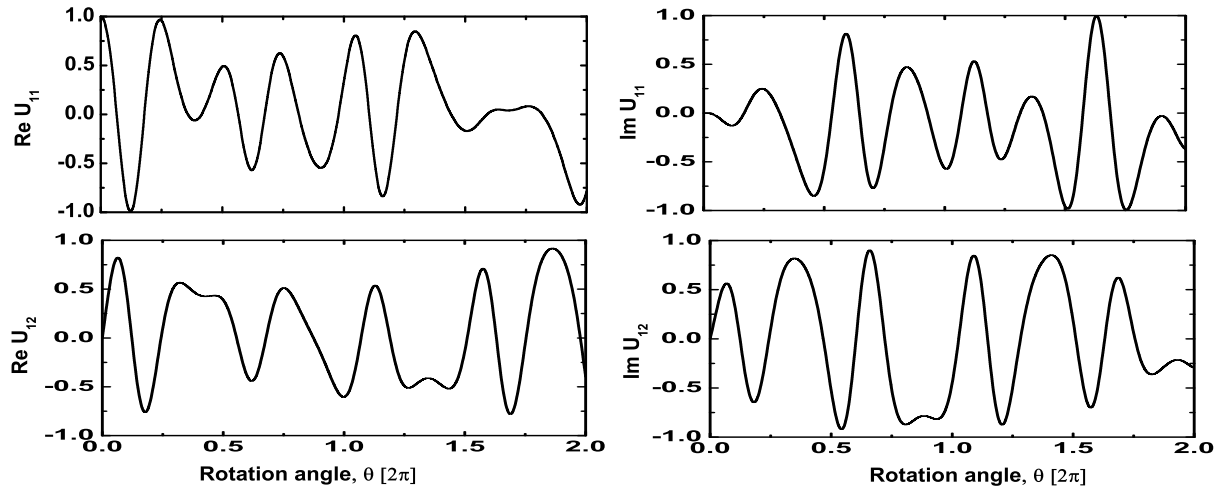}
\caption{\label{fig3}  Components of the evolution operator ~(see Eq.~\ref{U} in Appendix~\ref{appendix-B}) vs rotation angle.  The parameters are chosen  the same as in Fig.~\ref{fig2}. Superposition of these components of the evolution operator induces spin-echo in the expectation values of the Pauli spin matrices in Fig.~\ref{fig2}  during the adiabatic transport of the QDs in the plane of 2DEG.   }
\end{figure*}
\begin{figure}
\includegraphics[width=8.5cm,height=5cm]{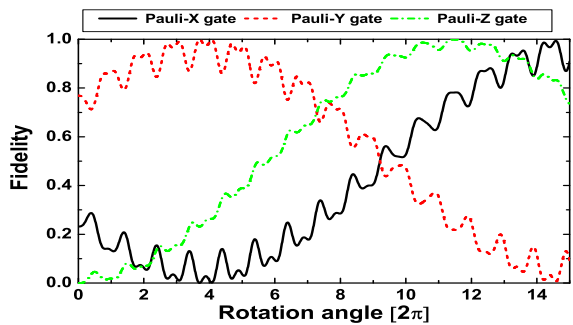}
\caption{\label{fig3a}   Evolution of three Pauli gate fidelities  during the adiabatic transport  of the dots in the 2D plane. Here the measurement of the  gate fidelity  is expressed in terms of the  probability between the objective or ideal vector state  and the  evolution of spin dynamics along the circular trajectory (see the text for details). The  parameters  are chosen the same as in Fig.~\ref{fig1}, but  $\ell_0=15$nm. }
\end{figure}
\begin{figure*}
\includegraphics[width=16cm,height=8cm]{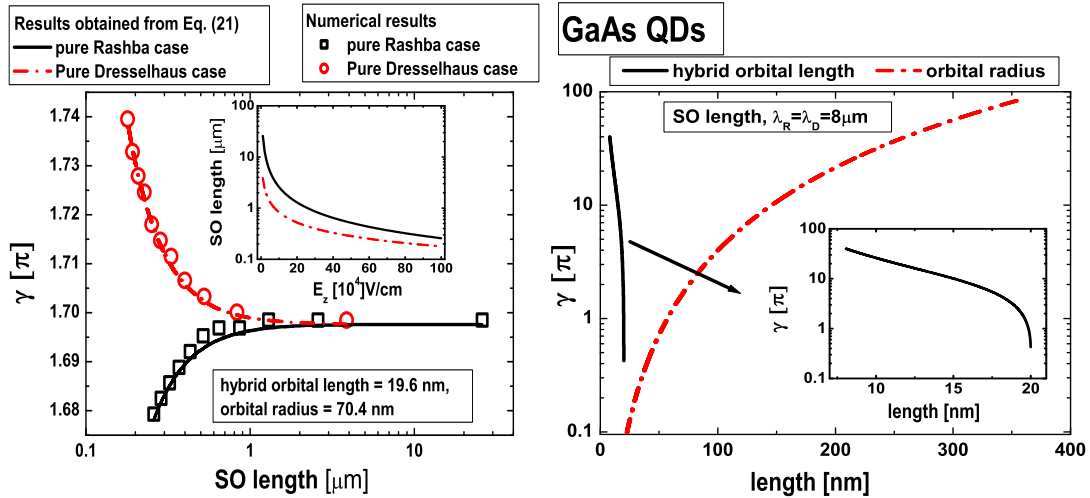}
\caption{\label{fig4} (Color online) (left panel) The Berry phase vs SO lengths on spin state $|0,0,+1\rangle$.
Here we chose $\ell_0=20~\mathrm{nm}$. (right panel) The Berry phase vs hybrid orbital length and orbital radius   on spin state $|0,0,+1\rangle$. Here we chose $\ell_0=20~\mathrm{nm}$ for solid line and $B=1~\mathrm{T}$ for dashed dotted line.
The parameters  are chosen the same as in Fig.~\ref{fig1}.   }
\end{figure*}
\begin{figure*}
\includegraphics[width=16cm,height=8cm]{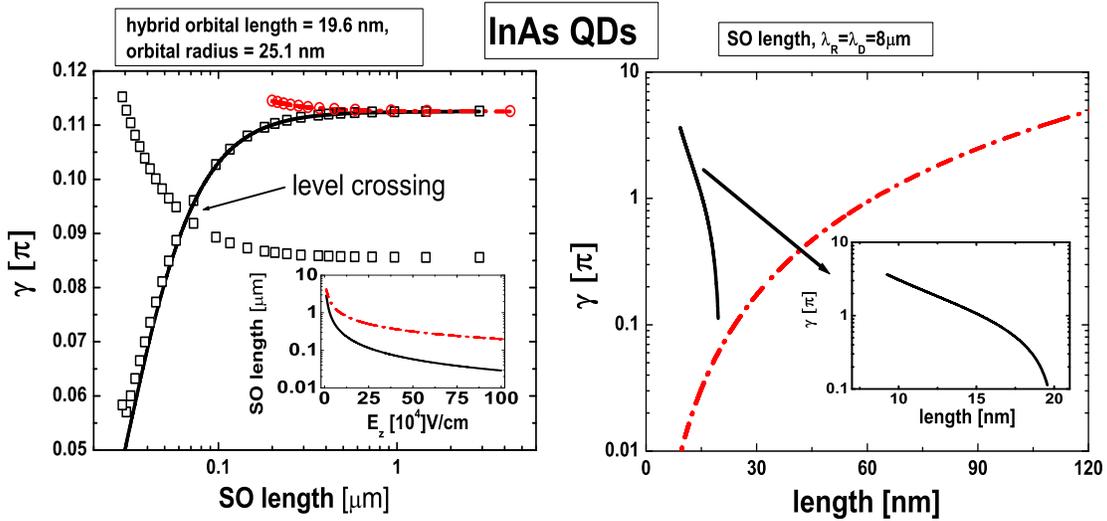}
\caption{\label{fig5} (Color online) See caption  to Fig.~\ref{fig4} (it is the same, but for InAs QDs).  }
\end{figure*}
\begin{figure*}
\includegraphics[width=16cm,height=8cm]{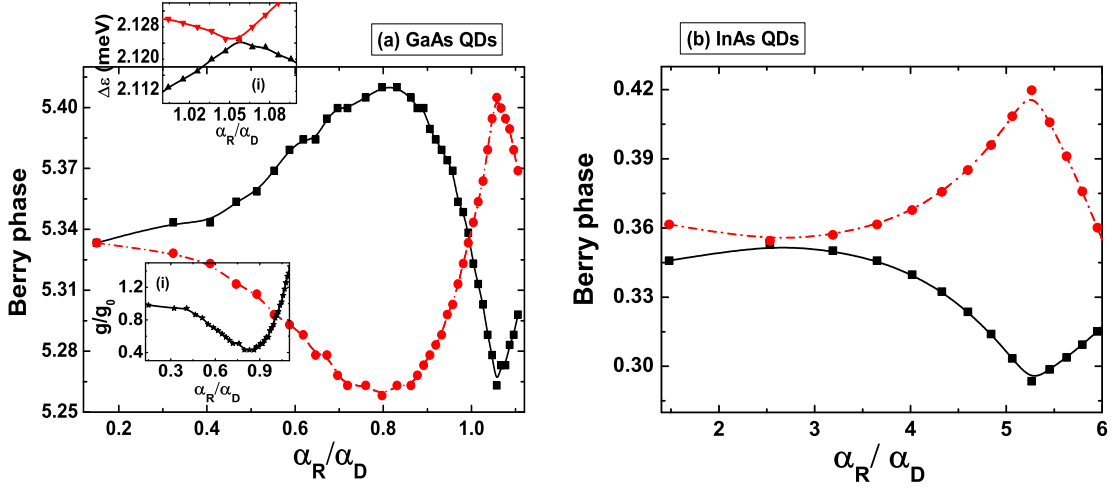}
\caption{\label{fig6} (Color online) The Berry phase (absolute value)  on the spin state $|0,0,+1\rangle$ (solid lines, filled circles) and  $|0,0,-1\rangle$ (dashed-dotted line, open circles) vs $\alpha_R/\alpha_D$ in  QDs. Inset plot (i) shows the energy difference ($\varepsilon_{0,1,+1}-\varepsilon_{0,0,-1}$ (triangle pointing down)) and ($\varepsilon_{0,1,-1}-\varepsilon_{0,0,-1}$ (triangle pointing up)) vs ratio of $\alpha_R$ to $\alpha_D$. Inset plot (ii) shows the variation of the $g$-factor $\left(g=\left(\varepsilon_{0,0,+1}-\varepsilon_{0,0,-1}\right)/\mu_B B\right)$ in QDs.  The parameters  are chosen the same as in Fig.~\ref{fig1}.  }
\end{figure*}
\begin{figure}
\includegraphics[width=8.5cm,height=6cm]{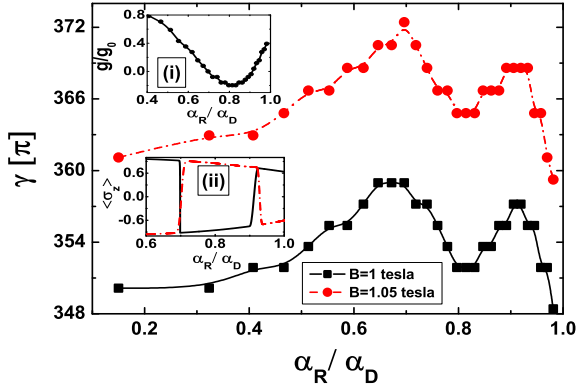}
\caption{\label{fig7} (Color online) The Berry phase (absolute value)  on the spin state $|0,0,+1\rangle$ vs $\alpha_R/\alpha_D$ in  GaAs QDs. The first maximum in the Berry phase can be seen due to the fact that the $g$-factor of electrons changes its sign (see inset plot (i)). The parameters  are chosen the  same as in Fig.~\ref{fig1}  but   $\ell_0=35$ nm and $B=1$T. Inset plot (ii) shows  $\langle\sigma_z\rangle$ for the states $|0,0,-1\rangle$ (solid line) and for the states $|0,0,+1\rangle$ (dashed-dotted line).}
\end{figure}
\begin{figure}
\includegraphics[width=8.5cm,height=6cm]{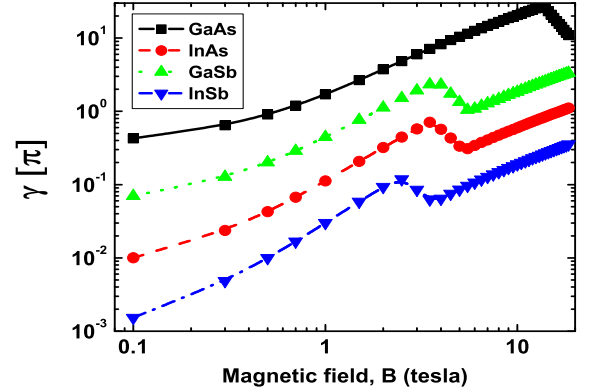}
\caption{\label{fig8} (Color online) The Berry phase (absolute value)  on the spin state $|0,0,+1\rangle$ vs longitudinal magnetic field  in  III-V semiconductor  QDs.  Here we choose $E_z=3\times 10^5~V/cm$ and   $\ell_0=20$ nm. Notice that the Berry phase terminates at the  crossing point.   }
\end{figure}

We now turn to the calculation of the Berry phase. If the QD is adiabatically carried around a circle of radius $r_0$, the wavefunction acquires a geometric phase,~\cite{berry84,wilczek84}
\begin{equation}
\gamma_n=-Im\int_S d\mathbf{S}  \mbox{\boldmath${\cdot}$}  \sum_{m\neq n}\frac{\langle n|  \mbox{\boldmath${\nabla_E}$}   \hat{H}|m \rangle  \mbox{\boldmath${\times}$} \langle m|\mbox{\boldmath${\nabla_E}$}  \hat{H}|n \rangle}{\left\{\varepsilon_m(\mathbf{E})-\varepsilon_n(\mathbf{E})\right\}^2},
\label{gamma-n}
\end{equation}
where  $S$ is the  area enclosed  by the circle. We consider $|n \rangle=|0,0,\pm 1  \rangle$ and $|m \rangle=|0,1,\pm 1  \rangle$ and investigate the Berry phase in QD accumulated during the adiabatic transport of the dot in the plane of 2DEG. Other choices of the parameters such as $|m \rangle=|1,0,\pm 1  \rangle$ also induce a non-zero Berry phase on $|n \rangle=|0,0,\pm 1 \rangle$  which is comparatively very small. Based on the second order perturbation theory, after lengthy algebraic transformations, Eq.~(\ref{gamma-n}) can be written as
\begin{equation}
\gamma_{0,0,\pm 1}= \mp \frac{\pi}{2}  \frac{\left(\hbar\omega_+\xi_-r_0\right)^2}{\left[\hbar\Omega_-\pm \xi_-^2\left(\frac{\alpha_R^2}{\varrho_+}-\frac{\alpha_D^2}{\varrho_-}\right)\right]^2},\label{gamma-n-1}
\end{equation}
where $\varrho_{\pm}=\hbar\Omega_-\pm\Delta$. Berry phase~(\ref{gamma-n-1}) can also be expressed in terms of three relevant length scales,  SO lengths ($\lambda_R$ and $\lambda_D$), hybrid orbital length ($\ell=\sqrt{\hbar/m\Omega}$) and orbital radius $(r_0=m e E_0 \ell^4_0/\hbar^2 )$, as:
\begin{equation}
\gamma_{0,0,\pm 1}= \mp \frac{\pi}{2}   \frac{\left(\ell \omega_+ r_0\right)^2 \left(2\hbar-eB\ell\right)^2 } {\left[\hbar\ell\Omega_- \pm \zeta \left(2\hbar-eB\ell\right)^2   \left( \lambda_D^2 \varrho_- -\lambda_R^2  \varrho_+   \right)\right]^2 },\label{gamma-n-2}
\end{equation}
where $\zeta=\hbar^2/2m^2\varrho_+\varrho_-\lambda_R^2\lambda_D^2$. The characteristics  of the Berry phase for three relevant  length scales (SO lengths,  hybrid orbital length and orbital radius) are discussed in Figs.~\ref{fig4} and~\ref{fig5}.

\section{Computational Method}\label{computational-method}

We suppose that a  QD is formed in the plane of a two dimensional electron gas of  $400\times 400~\mathrm{nm^2}$ geometry.  Then we vary the in-plane oscillating electric fields $E_x(t)$ and $E_y (t)$ adiabatically in such a way that the QD is transported in a closed loop of circular trajectory (see Fig.~\ref{fig1}). To find the Berry phase using an  explicit  numerical  method, we diagonalize the total Hamiltonian  $H(t)$ at any fixed time using the finite element method. The geometry contains  $24910$ elements.  Since the geometry is much larger compared to the actual lateral size of the QD,  we impose Dirichlet boundary conditions and  find the  eigenvalues and eigenfunctions  of the total Hamiltonian $H(t)$. In  Figs.~\ref{fig4} and~\ref{fig5}, the analytically obtained Berry phase from Eq.~(\ref{gamma-n-2}) (solid and dashed-dotted lines) is  seen to be in excellent agreement with the numerical values (circles and squares). Figs.~\ref{fig6}, \ref{fig7} and ~\ref{fig8} are obtained by solving the Hamiltonian $H(t)$ via the exact diagonalization method. The material constants for GaAs, InAs, GaSb and  InSb semiconductors are taken from Ref.~\onlinecite{prabhakar13}.

\section{Results and Discussions} \label{results-discussion}

In Fig.~\ref{fig1}, the realistic electron wavefunctions of the dots at  different locations ($\theta=0,\pi/2,\pi,3\pi/2$)  are shown. The evolution of spin dynamics in the expectation values of the Pauli spin matrices, due to adiabatic Rashba-Dresselhaus spin-orbit couplings in~(\ref{Had}), is shown in Fig.~\ref{fig2}. In the presence of both the Rashba and the Dresselhaus spin-orbit couplings, we find the spin-echo due to  a superposition of  spin waves in the evolution of spin dynamics during the adiabatic transport of the dots in the 2D plane (for details, see appendix~\ref{appendix-B} and Fig.~\ref{fig3}).

Since we know the exact unitary operator, it is possible to realize the quantum gates (see Fig.~\ref{fig3a}) during the adiabatic transport of the dots.~\cite{bednarek12,bednarek08,amparan13} In Fig.~\ref{fig3a}, we plot gate fidelity versus  rotation angle. Here we express the probability in terms of gate fidelity equal to $|\langle \Psi_{obj}|\chi\left(\theta\right)\rangle|^2$, where the objective  or ideal vector state $|\Psi_{obj}\rangle$ is  the product of the gate operation (Pauli matrix) on the initial state $|\chi \left(\theta=0\right)\rangle$ and $|\chi \left(\theta\right)\rangle$ is to evolve the dynamics of the unitary operator (see Eq.~\ref{chi}).~\cite{amparan13}
It can be seen that one can observe the perfect fidelity (i.e. fidelity=1) at $\theta=29\pi$ (solid line), $\theta=8\pi$ (dashed   line) and $\theta=23\pi$ (dashed-dotted line). Thus one can find the Pauli-X, Pauli-Y and Pauli-Z  gates at $\theta=29\pi, 8\pi$ and $23\pi$  respectively. Recently similar kind of results for the realization of Pauli gates from  symmetric graphene quantum dots by utilizing the genetic algorithm~\cite{chong01} have also been presented in Ref.~\onlinecite{amparan13}.

We now turn to  another key result of the paper: the analysis of the  Berry phase accumulated  during the adiabatic transport  of the dots in the 2D plane.
In Fig.~\ref{fig4}, we plot the characteristics of the Berry phase versus three relevant  length scales (SO length (left panel),   orbital radius and hybrid orbital length (right panel)). As can be seen (left panel of Fig.~\ref{fig4}), the Berry phase for the pure Rashba  and  pure Dresselhaus  cases  is  well separated at smaller values of the SO lengths  due to the presence of different symmetry orientations in the crystal lattice, such as  a lack of structural inversion asymmetry along the growth direction for the Rashba case and the bulk inversion asymmetry  for the Dresselhaus spin-orbit coupling case  [see Eq.~\ref{ratio}]. At large values of SO lengths $\lambda_R=\lambda > 1.8 \mu m$, the Berry phases for the pure Rashba and for the pure Dresselhaus spin-orbit coupling cases meet each other because the SO coupling strength is extremely weak and is unable to break the in-plane rotational symmetry.
Note that the SO length is characterized by the applied electric field along the z-direction (inset plot of the  left panel in Fig.~\ref{fig4} and also see Eqs.~(\ref{rashba}) and~(\ref{dresselhaus})).~\cite{prabhakar13} In the right panel of Fig.~\ref{fig4} (solid line), we see that the Berry phase decreases with increasing values of hybrid orbital length. This occurs because the hybrid orbital length is inversely proportional to the applied magnetic field  that reduce  the energy difference between the corresponding spin states. Also,  in the right panel of Fig.~\ref{fig4} (dashed dotted line), we see that the Berry phase increases with increasing values of orbital radius because of the enhancement in the total enclosed adiabatic area. Figure~\ref{fig5} investigates the characteristics of the Berry phase in InAs QDs with three relevant lengths: SO length, hybrid orbital length and QDs radii. For the pure Rashba case after the level crossing point at $\lambda_R=0.06~\mu m$, the analytically obtained values  from Eq.~(\ref{gamma-n-2}) capture the Berry phase on the state $|0,0,-1\rangle$.

In Fig.~\ref{fig6}(a), the abrupt changes (i.e.  the first maximum or minimum)   in  tunability of the Berry phase at  $\alpha_R/\alpha_D \approx 0.8$ are possible since  the Bloch wavefunctions can be pushed near the edge of   the barrier materials but are still located in the QD region because the effective $g$-factor of electrons   is still negative (see intet plot, Fig.~\ref{fig6}(a)(i)).
The second  maximum or minimum in the Berry phase at $\alpha_R/\alpha_D \approx 1.15$ can be seen due to the sign change in the $g$-factor of the p-state (see inset plot Fig.~\ref{fig6}(a)(ii)).
\label{RD}
In Fig.~\ref{fig6}(b), we study the Berry phase in InAs QDs.

Let us consider the quantitative difference between the Berry phases accumulated on the electron spin states $|0,0,\pm 1\rangle$. For simplicity, we only consider the second powers of the Rashba-Dresselhaus spin-orbit couplings:
\begin{equation}
\sqrt\frac{\gamma_{0,0,+1}}{\gamma_{0,0,-1}}=1-\frac{2m}{\hbar^3\Omega}\left[\alpha_R^2\left(1-\frac{\Delta}{\hbar\Omega_-}\right)
-\alpha_D^2\left(1+\frac{\Delta}{\hbar\Omega_-}\right)\right].
\label{ratio}
\end{equation}
In InAs and InSb QDs, $\alpha_R > \alpha_D$. It means,  $\gamma_{0,0,-1}>\gamma_{0,0,+1}$ and viceversa for  GaAs and GaSb QDs.

In Fig.~\ref{fig7},  we find that the large enhancement in the Berry phase occurs  with a very small increment in the magnetic fields. This indicates  that the Berry phase is  highly sensitive to magnetic fields in QDs. The first maximum (approx. $\alpha_R/\alpha_D=0.67$) in the Berry phase results from the sign change in the $g$-factor of electrons in QDs (see the inset plot of Fig.8). This means that  the Bloch wavefunctions start  penetrating into the barrier materials. Experimentally, the penetration of the Bloch wavefunctions in  the AlGaAs/GaAs heterojunction can be engineered with the application of  gate controlled electric fields along the z direction where the bulk $g$-factor of electrons for GaAs materials is negative, and for AlGaAs it is positive.~\cite{jiang01} The second maximum  (at $\alpha_R/\alpha_D=0.9$) can be seen due to the fact that the wavefunctions of electrons are pushed  back  into the GaAs material. The minimum point (at  $\alpha_R/\alpha_D=0.9$)  in  the Berry phase and in the $g$-factor indicates that the Bloch wavefunctions are getting pushed
towards the QD region due to the interplay between the Rashba and the Dresselhaus spin-orbit couplings.
Fig.~\ref{fig8} investigates the Berry phase versus  magnetic fields in III-V semiconductor QDs.

\begin{figure}
\includegraphics[width=8.5cm,height=7cm]{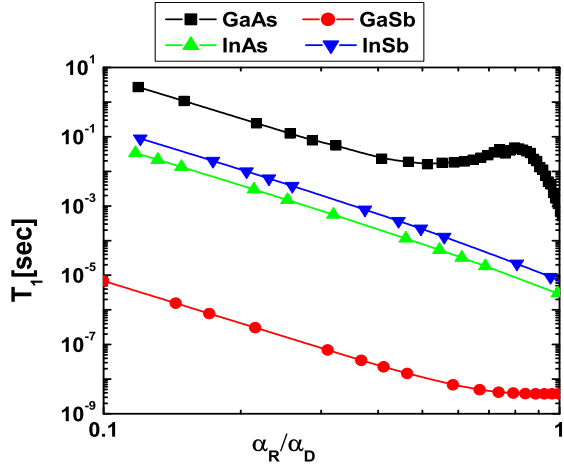}
\caption{\label{fig8a}
(Color online) Spin relaxation time, $T_1$ vs $\alpha_R/\alpha_D$ in a moving QD at location $\theta=0$ (see Fig.~\ref{fig1}). We choose   $B=1$ T, $\ell_0=20$ nm and $E_0=5\times 10^3$ V/cm. The material constants are chosen from Ref.~\onlinecite{sousa03}.
}
\end{figure}

\textbf{Spin relaxation:} Now we estimate the spin relaxation time caused by the emission of one phonon  at absolute zero temperature between two lowest energy states in III-V semiconductor QDs. Since we deal with small energy transfer between electron in QDs and phonon, we only consider a piezo-phonon.\cite{khaetskii01} Hence  coupling between  an electron and a piezo-phonon with mode $\mathbf{q} \alpha$ ($\mathbf{q}$ is the phonon wave vector and the branch index $\alpha=l,t_1,t_2$ for one longitudinal  and two transverse modes) is given by:~\cite{khaetskii01,sousa03,prabhakar13}
\begin{equation}
u^{\mathbf{q}\alpha}_{\mathrm{ph}}\left(\mathbf{r},t\right)=\sqrt{\frac{\hbar}{2\rho V \omega_{\mathbf{q}\alpha}}} e^{i\left(\mathbf{q\cdot r} -\omega_{q\alpha} t\right)}e A_{\mathbf{q}\alpha}b^{\dag}_{\mathbf{q}\alpha} + \mathrm{H.c.},
\label{u}
\end{equation}
where  $\rho$ is the crystal mass density, $V$ is the volume of the QD and   $A_{\mathbf{q}\alpha}=\hat{q}_i\hat{q}_k e\beta_{ijk} e^j_{\mathbf{q}\alpha}$ is the amplitude of the electric field created by phonon strain. Here $\hat{\mathbf{q}}=\mathbf{q}/q$ and $e\beta_{ijk}=eh_{14}$ for $i\neq k, i\neq j, j\neq k$.  Based on the Fermi Golden Rule, the phonon induced spin transition rate in the QDs is given by~\cite{sousa03,khaetskii01}
\begin{equation}
\frac{1}{T_1}=\frac{2\pi}{\hbar}\int \frac{d^3\mathbf{q}}{\left(2\pi\right)^3}\sum_{\alpha=l,t}\arrowvert M\left(\mathbf{q}\alpha\right)\arrowvert^2\delta\left(\hbar s_\alpha \mathbf{q}-\varepsilon_{f}+\varepsilon_{i}\right),
\label{1-T1}
\end{equation}
The matrix element $M\left(\mathbf{q}\alpha\right)=\langle \psi_i|u^{\mathbf{q}\alpha}_{ph}\left(\mathbf{r},t\right)|\psi_f\rangle$   has been calculated numerically.~\cite{prabhakar13,comsol} Here $|\psi_i\rangle$ and $|\psi_f\rangle$ correspond to the initial and finial states of the Hamiltonian $H$.

In Fig.~\ref{fig8a}, we plotted the spin relaxation time versus the interplay between the Rashba and Dresselhaus spin-orbit coupling strengths. Different behavior of spin-relaxation in GaAs is observed due to the fact that the $g$-factor of electron spin in GaAs QDs changes its sign (see the inset plot of Fig.~\ref{fig7}). Evidently large spin relaxation time, $T_1$ and thus decoherence time, $T_2 \approx 2T_1$ can be seen in GaAs QDs.

Long decoherence time  combined with  short gate operation time is one of the requirements for quantum computing and quantum information processing.~\cite{loss98,awschalom02,bandyopadhyay00}
However, at (or near by) the level crossing point  in the Berry phase, a spin-hot spot can be observed that greatly reduces  the decoherence time.~\cite{bulaev05,bulaev05a,sousa03,prabhakar13} Thus one should avoid such level crossing points in the Berry phase during the design of QD spin-based transistors for possible implementation in  solid-state quantum computing and quantum information processing.
When a qubit is operated on by a classical bit, then its decay time is given by a spin-relaxation time which is also supposed to be longer than the minimum time required to execute one quantum gate operation.~\cite{amasha08,pribiag13} It seems that the spin-relaxation time in GaAs QD is much larger than in other materials (InAs, InSb and GaSb) due to the presence of weak spin-orbit coupling.~\cite{prabhakar13,bulaev05,bulaev05a} However, other factors such as mobility of the charge carriers and defects might greatly affect the performance of gate operation time, and hence decoherence time. Thus, additional experimental studies may be required to show that GaAs is indeed a better candidate for quantum gate operations.
Enhancement in the Berry phase of GaAs QDs and  extension of the level crossing point, such as to larger QDs radii as well as to   larger magnetic fields, might provide some additional benefits to control  electron spins for larger lateral size  QDs when choosing GaAs material rather than InAs, InSb, or GaSb.

Finally, we mention a possible experimental realization of the measurement of the  Berry phase in QDs. Several parameters such as $E_x(t)$, $E_y(t)$ and $\theta$ in the distortion potential can relate to the other control parameters,  $\alpha_R$, $\alpha_D$, $\omega_0$, and $\Delta/\hbar$  of the dots, so that    one can experimentally realize the  adiabatic  movement of the QDs in the 2D plane. Following Refs.~\onlinecite{berger12,pechal12,hwang06,bason12}, the adiabatic movement of the dots can be performed by choosing the frequency $\omega$ of the microwave pulse smaller than $\varepsilon^0_{0,0,\pm 1}/\hbar$ and $\omega_0$. Also, we chose $E_0 \ll E_z$ to study the  interplay between the Rashba and the Dresselhaus spin-orbit couplings on the Berry phase.

\section{Conclusion}\label{conclusion}

We have calculated the evolution of the spin dynamics and the superposition due to the Rashba-Dresselhaus spin-orbit couplings that can be seen during the adiabatic transport  of QDs in the 2D plane.  We have shown  that  the Berry phase in the lowest Landau levels of the QD can be generated from  higher quantum states that  only differ by one quantum number of the corresponding spin states. The Berry phase is  highly sensitive to the magnetic fields, QD radii, and the Rashba-Dresselhaus spin-orbit coupling coefficients.  We have shown that the sign change in the $g$-factor  in the Berry phase can be manipulated with the interplay between the Rashba and the Dresselhaus spin-orbit couplings. We have provided a detailed analysis of the characteristics of the Berry phase with three relevant length scales (SO length, hybrid orbital length and orbital  radius). The sets of data, which can be encoded at the degenerate sub-levels (i.e. at $g=0$) but well separated in their phase,   are topologically protected and can help to  build  a topological solid-state quantum computer.

\section{Acknowledgements:} This work was supported by NSERC and CRC programs, Canada. The authors acknowledge Prof. Akira Inomata from the State University of New York at Albany for his many helpful discussions. The authors also acknowledge the Shared Hierarchical Academic Research Computing Network (SHARCNET) community  and Dr. P.J.  Douglas Roberts for his assistance and technical support.

\appendix

\section{ Induced magnetic field due to oscillating electric field}\label{appendix-A}

Induced displacement current density due to oscillating electric field is given by
\begin{equation}
\mathbf{J_{in}}=\epsilon_0\epsilon_r \frac{\partial \mathbf{E}(\mathbf{r})}{\partial t}=\frac{\epsilon_0\epsilon_r\omega E_0}{r_0}\left(y_0,-x_0\right),
\end{equation}
where $x_0=-r_0 \cos \omega t$, $y_0=-r_0 \sin \omega t$, $\epsilon_r$ is the relative permittivity  and  $\epsilon_0$ is the permittivity of the free space. The induced current due to $\mathbf{J_{in}}$ is
approximated as
\begin{equation}
I_{in}=\pi r_0^2|\mathbf{J_{in}}|=\pi r_0^2\epsilon_0\epsilon_r\omega E_0. \label{Iin}
\end{equation}
We apply Ampere's law to estimate the induced $B$ field at the center of the orbit:
\begin{equation}
B_z=\frac{\mu_0\mu_r I_{in}}{2r_0}=\frac{\pi\epsilon_r\mu_r r_0 \omega E_0}{2c^2},
\end{equation}
where $\mu_r$ is the relative permeability and $c=1/\sqrt{\epsilon_0\mu_0}$ is the velocity of light with $\mu_0$ being the permeability of the free space.

\section{ Exact unitary operator of spin Hamiltonian} \label{appendix-B}
\begin{figure}
\includegraphics[width=9cm,height=5.5cm]{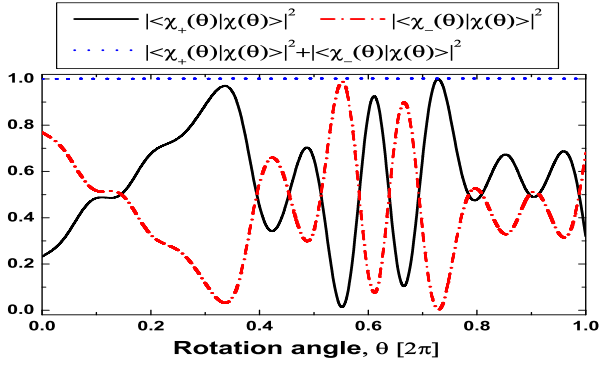}
\caption{\label{fig9} (Color online) The exact probability of a transition to spin up (solid line) and spin down (dashed-dotted line) vs rotation angle.      The  parameters  are chosen the same as in Fig.~\ref{fig2}.   }
\end{figure}

To investigate the evolution of spin dynamics due to adiabatic parameters in the Hamiltonian~((\ref{Had}), we write the adiabatic Rashba-Dresselhaus spin-orbit couplings as
\begin{equation}
h_{ad}=\frac{1}{\hbar}\left(\alpha_Rz_--i\alpha_Dz_+\right)m\omega_+s_+ +H.c.,\label{Had-1}
\end{equation}
where  $s_{\pm}=s_x\pm  i s_y$. We construct a normalized orthogonal set of eigenspinors of Hamiltonian~(\ref{Had-1}) as:
\begin{eqnarray}
\mathbf{\chi_+}\left(t\right)=\frac{1}{\sqrt 2}
\left(\begin{array}{c}
1 \\
\frac{\left[\alpha_1^2+\beta_1^2+2\alpha_1\beta_1\sin 2\omega t\right]^{1/2}}{i\beta_1\exp\left(i\omega t\right)-\alpha_1 \exp\left(-i\omega t\right)} \\
\end{array}\right),\label{chi-p-1}\\
~\nonumber\\
\mathbf{\chi_-}\left(t\right)=\frac{1}{\sqrt 2}
\left(\begin{array}{c}
\frac{i\beta_1\exp\left(i\omega t\right)-\alpha_1 \exp\left(-i\omega t\right)}{\left[\alpha_1^2+\beta_1^2+2\alpha_1\beta_1\sin 2\omega t\right]^{1/2}} \\
-1 \\
\end{array}\right),\label{chi-m-1}
\end{eqnarray}
where $\alpha_1=\alpha_R r_0 m \omega_+/\hbar$, $\beta_1=\alpha_D r_0 m \omega_+/\hbar$. Following Ref.~(\onlinecite{prabhakar10}), by utilizing the disentangling operator technique,  the exact evolution operator of~(\ref{Had-1}) for a spin-1/2 particle can be written as:
\begin{eqnarray}
U(t)&=& \tau \exp\left\{ \frac{-i}{\hbar} \int h_{ad} ~d \tau  \right\},\label{U-1}\\
&=&\left(\begin{array}{cc}\exp\left\{{\frac{b}{2}}\right\} + ac\exp\left\{{-\frac{b}{2}}\right\} & ~~~a\exp\left\{-{\frac{b}{2}}\right\}\\
c\exp\left\{-{\frac{b}{2}}\right\} & ~~~\exp\left\{-{\frac{b}{2}}\right\} \end{array}\right),~~~~~ \label{U}
\end{eqnarray}
where $\tau$ is a time ordering parameter. The components of the  evolution operator follow the relation: $U_{22}=\mathrm{conj}(U_{11})$ and $U_{12}=-\mathrm{conj}(U_{21})$. The $\theta$ dependent functions $a(\theta)$, $b(\theta)$, and $c(\theta)$ are written in terms of adiabatic control parameters $x_0$ and $y_0$ as:
\begin{eqnarray}
\frac{d a}{d\theta}= \frac{i}{\hbar^2} m \varpi_+ \{\left(-\alpha_-x_0 +i \alpha_+ y_0\right)\nonumber\\
+  a^2 \left(\alpha_+ x_0 + i \alpha_- y_0\right)\}, \label{a}\\
\frac{d b}{d\theta}= \frac{2i}{\hbar^2} m \varpi_+ \left(\alpha_+ x_0 + i \alpha_- y_0\right) a, \label{b}\\
\frac{d c}{d\theta}= -\frac{i}{\hbar^2} m \varpi_+  \left(\alpha_+ x_0 + i \alpha_- y_0\right) e^b, \label{c}
\end{eqnarray}
where $\alpha_{\pm}=\alpha_R \pm i \alpha_D$ and $\varpi_+=1+\omega_c/(2\omega)$. At $\theta=0$, we use the initial condition $\chi\left(0\right)=\left(1~\sqrt{\alpha_1^2+\beta_1^2}/\left(i\beta_1-\alpha_1\right)\right)^\top/\sqrt 2$, where $\top$ denotes transpose and write $\chi\left(t\right)=U(t,0)\chi\left(0\right)$ as
\begin{equation}
\chi\left(t\right)=\frac{1}{\sqrt 2}\left(\begin{array}{c} e^{b/2} + a c e^{-b/2} + \left(\frac{\sqrt{\alpha_1^2+\beta_1^2}}{i\beta_1-\alpha_1}\right) a e^{-b/2} \\ \nonumber
ce^{-b/2} +  \left(\frac{\sqrt{\alpha_1^2+\beta_1^2}}{i\beta_1-\alpha_1}\right) e^{-b/2} \end{array}\right). \label{chi}
\end{equation}
By using Eq.~(\ref{chi}), we found  expectation values of  Pauli spin matrices and plotted them in Fig.~\ref{fig2}. Components of the evolution operator of~(\ref{U}) are shown in Fig.~\ref{fig3}.
The exact probabilities of a transition to spin up (solid line) and spin down (dashed-dotted line) are shown in Fig.~\ref{fig9} during the adiabatic movement of the QDs in the 2D plane. It can be seen that the sum of spin up and spin down probabilities  are always  unity (dotted line of Fig.~\ref{fig9}) which indicates that the evolution operator~(\ref{U}) of the quasi-Hamiltonian~(\ref{Had-1}) is exact and the symmetry of the unitary operator during the adiabatic movement of dots is preserved.

In order to verify that the evolution operator~(\ref{U}) is indeed exact and unitary to~(\ref{U-1}),  we expand~(\ref{U-1}) for the pure Rashba case by following Dyson series method as:
\begin{widetext}
\begin{eqnarray}
U(t)&=&1+i \tilde{r}_0 \left\{ \sigma_x \sin\theta +\sigma_y \left( 1-\cos\theta \right) \right\}  - \tilde{r}^2_0 \left\{ -i \sigma_z \left( \theta -\sin\theta \right) + \left(1-\cos\theta\right) \right\} \nonumber \\
&&- i \tilde{r}^3_0 \left\{  \sigma_x \left( 2\sin\theta -\theta\cos\theta -\theta \right) + \sigma_y \left( -2\cos\theta -\theta \sin\theta + 2 \right) \right\} \nonumber \\
&&+ \tilde{r}^4_0 \left\{ i \sigma_z \left(-2\theta +3\sin\theta -\theta\cos\theta \right) + \left( -\frac{1}{2} \theta^2 -3\cos\theta -\theta\sin\theta +3  \right) \right\}
+O(\tilde{r}^5_0), \label{U-3}
\end{eqnarray}
\end{widetext}
where $\tilde{r}_0=\alpha_R m\varpi_+ r_0/\hbar^2$. Next we write evolution operator~(\ref{U}) as
\begin{equation}
U(t)=U_0 I + U_x \sigma_x + U_y \sigma_y + U_z \sigma_z,\label{U-2}
\end{equation}
where
\begin{eqnarray}
U_0=\frac{1}{2} \left\{ \exp\left( \frac{b}{2} \right) + a c \exp\left( \frac{b}{2} \right) +  \exp\left( -\frac{b}{2} \right)    \right\},~~~~~~~~\label{U0}\\
U_x=\frac{1}{2} \left\{  a \exp\left(- \frac{b}{2} \right) + c  \exp\left( -\frac{b}{2} \right)    \right\},  ~~~~~~~~~ \label{Ux}\\
U_y=\frac{i}{2} \left\{  a \exp\left(- \frac{b}{2} \right) - c  \exp\left( -\frac{b}{2} \right)    \right\},   ~~~~~~~~~ \label{Uy}\\
U_z=\frac{1}{2} \left\{ \exp\left( \frac{b}{2} \right) + a c \exp\left( -\frac{b}{2} \right) -  \exp\left( -\frac{b}{2} \right)    \right\}.  ~~~~~~~~\label{Uz}
\end{eqnarray}
The functions $a(\theta)$, $b(\theta)$,  $c(\theta)$ are obtained by solving three coupled Riccatti Eqs.~(\ref{a}), (\ref{b}),  (\ref{c}) for the pure Rashba case as
\begin{eqnarray}
a(\theta)= \frac{ 2 \tilde{r}_0 \left\{ \exp (-i n_1 \theta)  -\exp\left(-i\theta\right)\right\}  }{ n_2 - n_1 \exp \left\{-i \left( n_1- 1\right) \theta \right\} }, \label{a-theta}\\
\exp (b(\theta)/2)= \frac{ 2 \left( n_1 -1 \right)  \exp (-i n_1 \theta/2)   }{  n_1 \exp \left\{-i \left( n_1- 1\right) \theta \right\} - n_2 }, \label{b-theta}\\
c(\theta)= \frac{ 2 \tilde{r}_0 \left\{ 1- \exp \left[-i\left( n_1-1\right) \theta\right] \right\}  }{ n_1 \exp \left\{-i \left( n_1- 1\right) \theta \right\} - n_2 }, \label{c-theta}
\end{eqnarray}
where
\begin{eqnarray}
n_{1,2}=1 \pm \sqrt{1+4 \tilde{r}^2_0}.
\end{eqnarray}
By substituting Eqs.~(\ref{a-theta}), (\ref{b-theta}) and  (\ref{c-theta}) in Eqs.~(\ref{U0}), (\ref{Ux}), (\ref{Uy}) and  (\ref{Uz}), we find
\begin{widetext}
\begin{eqnarray}
U_0= 1+  2 \frac{\tilde{r}^2_0}{2!} \left(\cos\theta-1\right) + 24 \frac{\tilde{r}^4_0}{4!} \left(3-\frac{1}{2} \theta^2 -\theta\sin\theta-3\cos\theta  \right) + O\left(\tilde{r}^6_0\right) ,\label{U0-1}\\
U_x= i \tilde{r}_0 \sin\theta +  6 i \frac{\tilde{r}^3_0}{3!} \left(-2\sin\theta+\theta\cos\theta + \theta \right) + O\left(\tilde{r}^5_0\right)    ,   \label{Ux-1}\\
U_y= i \tilde{r}_0 \left( 1-\cos\theta\right) +  6 i \frac{\tilde{r}^3_0}{3!} \left(-2 + 2\cos\theta+\theta\sin\theta \right) + O\left(\tilde{r}^5_0\right)  ,    \label{Uy-1}\\
U_z=   2 i \frac{\tilde{r}^2_0}{2!} \left(\theta-\sin\theta \right) + 24 i \frac{\tilde{r}^4_0}{4!} \left(-2\theta -\theta\cos\theta+3\sin\theta \right) + O\left(\tilde{r}^5_0\right) .  \label{Uz-1}
\end{eqnarray}
\end{widetext}
One can easily identify that the coefficients $U_0$, $U_x$, $U_y$ and  $U_z$  of Eq.~(\ref{U-2}) are exactly the same as in Eq.~(\ref{U-3}). Thus the evolution operator~(\ref{U}) obtained by the Feynman disentangling operator scheme is exact  for any order of the orbital radius.


%

\end{document}